\input{aipcheck}

\documentclass[final,sort&compress]{aipproc}
\usepackage{amsfonts}
\usepackage{amsmath}
\usepackage{amssymb}
\usepackage{graphicx}

\layoutstyle{6x9}

\begin{document}

\title{Nonextensive statistical mechanics and central limit theorems I - Convolution of independent random variables and $q$-product}


\author{Constantino~Tsallis}{
  address={Centro Brasileiro de Pesquisas F\'{\i}sicas, 150, 22290-180, Rio de Janeiro   
- RJ, Brazil}, altaddress={Santa Fe Institute, 1399 Hyde Park Road, Santa Fe, New Mexico 87501, USA}
}

\author{S\'{\i}lvio M. Duarte~Queir\'{o}s}{
  address={Centro Brasileiro de Pesquisas F\'{\i}sicas, 150, 22290-180, Rio de Janeiro
- RJ, Brazil}
}

\begin{abstract}

In this article we review the standard versions of the Central and of the L\'evy-Gnedenko Limit Theorems,
and illustrate their application to the convolution of independent random variables
associated with the distribution $\mathcal{G}_{q}\left( X\right) \equiv 
\mathcal{A}_{q}\left[ 1+\left( q-1\right) \mathcal{B}_{q}\left( X-\bar{\mu}%
_{q}\right) ^{2}\right] ^{\frac{1}{1-q}}$ $(\mathcal{A}_{q}>0; \,\mathcal{B}_{q}>0;\, q<3$), known as $q$-Gaussian. This
distribution emerges upon extremisation of the nonadditive entropy $S_{q}\equiv
k\left( 1-\int \left[ p\left( X\right) \right] ^{q}\ dX\right) /\left(
1-q\right) $, basis of nonextensive statistical
mechanics. It has a finite variance for $q<5/3$, and an infinite one for $q \ge 5/3$. We exhibit that, in the case of (standard) independence, the $q$-Gaussian
has either the Gaussian (if $q<\frac{5}{3}$) or the $\alpha $-stable L\'{e}%
vy distributions (if $q > \frac{5}{3}$) as its attractor in probability
space. Moreover, we review a generalisation of the product, the $q$-product,
which plays a central role in the approach of the specially correlated variables emerging within the nonextensive theory.

\end{abstract}

\maketitle

\section{Introduction}

\label{intro}

Science, whether in its pure or applied form, is frequently related to the
description of the behaviour exhibited by systems when some quantity
approaches a particular value. We might consider the effect on the state of a
particle as we furnish it a certain energy, $E$, when this quantity tends to some
critical value, $E_{c}$, or the response of a system when the number of
elements goes to infinity, like it usually occurs in  thermodynamics and
statistical mechanics, i.e., the \textit{thermodynamic limit}. In the latter case, we
may focus on the outcome of the addition, or arithmetic average, of a large (infinite) sequence of
random variables associated with a certain observable. This constitutes the
basis of the celebrated central limit theorem (CLT), which is at the core of the theory of
probabilities and mathematical statistics. The CLT has its origin at the weak
law of large numbers of \textsc{Jacob Bernoulli} \cite{bernoulli}. For
independent random variables, it had its first version introduced by 
\textsc{Abraham de Moivre} in $1733$ \cite{moivre}, who used the normal
distribution to approximate the functional form of a binomial distribution
for a large number of events. In $1812$, his result was later extended by 
\textsc{Pierre-Simon Laplace}, who formulated the now called \textit{Theorem
of de Moivre-Laplace} \cite{laplace}. Laplace also used the normal distribution
in the analysis of errors in experiments, but it was \textsc{Carl Friedrich
Gauss} who first proved, in $1809$, the connection between error in
measurement and the normal distribution. It is due to this relation that the
normal distribution is widely called in Physics as \emph{Gaussian
distribution}. Although discovered more than once by different people, the
fact is that only in $1901$, mathematician \textsc{Aleksandr Lyapunov}
defined the CLT in general terms and proved it in a precisely mathematical
fashion \cite{lyapunov,tijms}.

After the establishment of a central limit theorem for the addition of
independent random variables with finite second-order moment, other versions
have appeared, namely, the \textit{L\'{e}vy-Gnedenko extension} for the
sum of independent random variables with diverging second-order moment \cite%
{levy,gnedenko}, the \textit{m-dependent central limit theorem}, \textit{%
martingale central limit theorem} \cite{hall}, the \textit{central limit
theorem for mixing processes} among others \cite%
{artstein,ct-quimicanova,ct-gm,schlogel,ct-gm-sato,goyo-ct-gm,q-clt-gauss,q-clt-levy1,q-clt-levy2,araujo}%
.

In this article, we review the fundamental properties of nonadditive
entropy, $S_{q}$, its optimising distribution (known as $q$-Gaussian), and the $q$-product, a
generalisation of the product \cite{nivanen,borges} formulated within
nonextensive statistical mechanics. We analyse, both analytically and
numerically, the sum of conventional independent random variables and show
that, in this case, the attractor in \ probability space is the Gaussian
distribution if random variables have a finite second-order moment, or the $%
\alpha $-stable L\'{e}vy distributions otherwise.

\section{Nonadditive entropy $S_{q}$}

Statistical mechanics, \textit{i.e.}, the application of statistics to large
populations whose state is governed by some Hamiltonian functional, is
strongly attached to the concept of entropy originally introduced by \textsc{%
Rudolf Julius Emmanuel Clausius} in $1865$~\cite{fermi}. The relation
between entropy and the number of allowed microscopic states was firstly
established by \textsc{Ludwig Eduard Boltzmann} in $1877$ when he was studying the
approach to equilibrium of an ideal gas~\cite{boltz}. Mathematically, this
relation is,%
\begin{equation}
S=k\,\ln W,  \label{boltz-prin}
\end{equation}%
where $k$ is a positive constant and $W$ the number of microstates
compatible with the macroscopic state. This equation is known as \emph{%
Boltzmann principle}.

When a system is not isolated, but instead in contact with some kind of
reservoir, it is possible to derive, from Eq.~(\ref{boltz-prin}) under some
assumptions, the Boltzmann-Gibbs entropy, $S_{BG}=-k\sum%
\limits_{i=1}^{W}p_{i}\,\ln p_{i}$, where $p_{i}$ is the probability of
microscopic configuration $i$~\cite{SBG}. Boltzmann-Gibbs
statistical mechanics is based on the molecular chaos~\cite{boltz} and
ergodic~\cite{khinchin} hypotheses \cite{cohen}. It has been very
successful in the treatment of systems in which {\it short} spatio/temporal
interactions dominate. In this case, ergodicity and independence are
justified and Khinchin's approach to $S_{BG}$ is valid~\cite{khinchin}.
Therefore, it appears as entirely plausible that physical entropies other than the
Boltzmann-Gibbs one, can be defined in order to treat anomalous systems, for
which ergodicity and/or independence are not verified.

Inspired by this kind of systems it was proposed in $1988$ \cite{ct88} the
entropy~$S_{q}\equiv k\left( 1-\sum\limits_{i=1}^{W}p_{i}^{q}\right) /\left(
1-q\right) $ ($q\in \Re ; \,
\lim_{q\rightarrow 1}S_{q}=S_{BG}$) as the basis of a possible extension of Boltzmann-Gibbs statistical mechanics 
\cite{further,3-gen} where the \textit{entropic index }$q$ should be
determined \textit{a priori} from microscopic dynamics. Just like $S_{BG}$, $%
S_{q}$ is \textit{nonnegative}, \textit{concave} ($\forall {q>0}$), \textit{experimentally
robust} (or \textit{Lesche-stable}~\cite{props}) ($\forall {q>0}$), composable, and
leads to a \textit{finite entropy production per unit time}~\cite%
{qprops,latorabaranger99}. Moreover, it has been shown~ that it is also 
\textit{extensive} \cite{ct-gm-sato,additive,marsh,marsh-moyano}, hence in
compliance with Clausius concept on macroscopic entropy and thermodynamics,
for a special class of \textit{correlated} systems. More precisely, systems whose phase-space is occupied in a (asymptotically) scale-invariant manner. It is upon this kind
of correlations that the $q$-generalised Central Limit Theorems are constructed.

At this stage let us emphasize the difference between the \textit{%
additivity} and \textit{extensivity} concepts for entropy \footnote{%
In this discussion we will treat elements of a system as strictly identical and distinguishable.}%
. An entropy is said to be \emph{additive} if \cite{penrose} if for two \emph{%
probabilistically independent} systems, let us say $A$ and $B$, the total
entropy equals the sum of the entropies for the two independent systems, \textit{%
i.e.}, $S\left( A+B\right) =S\left( A\right) +S\left( B\right) $. According
to this definition, Boltzmann-Gibbs entropy, $S_{BG}$, and R\'{e}nyi
entropy, $S_{\alpha }^{R}$ \cite{renyi}, $S_{\alpha }^{R}=\frac{1}{1-\alpha }%
\ln \left[ \sum_{i=1}^{n}p_{i}^{\alpha }\right] $, are \textit{additive},
while $S_{q}$ ($q\neq 1$), among others \cite{ct-gm}, is \textit{nonadditive}%
. Despite the fact of being nonadditive, $S_{q}$, just as additive entropies 
$S_{BG}$ and $S_{\alpha }^{R}$, is \textit{composable}, as already mentioned. By this we mean
that, for a system composed by two independent subsystems, $A$ and $B$, if we know the
entropy of each sub-system, then we are able to evaluate the entropy of the
entire system. Composability for $S_{BG}$ and $S_{\alpha }^{R}$ is a
consequence of its additivity, whereas for $S_{q}$ it results from the fact
that, considering independent subsystems, the total entropy satisfies%
\begin{equation}
\frac{S_{q}\left( A+B\right) }{k}=\frac{S_{q}\left( A\right) }{k}+\frac{%
S_{q}\left( B\right) }{k}+\left( 1-q\right) \frac{S_{q}\left( A\right) }{k}%
\frac{S_{q}\left( B\right) }{k} \,.  \label{nonadditivity}
\end{equation}%
On the other hand, an entropy is defined as \emph{extensive }whenever the
condition,%
\begin{equation}
\lim_{N\rightarrow \infty }\frac{S(N)}{N}=s \in (0,\infty) ,
\label{extensive-entropy}
\end{equation}%
is verified ($N$ represents the number of elements of the system). In
this definition, the correlation between elements of the system is arbitrary, 
\textit{i.e.}, \textit{it is not important to state whether they are
independent or not}. 
If the elements of the system are independent (e.g., the ideal gas, the ideal paramagnet), then the additive entropies $S_{1}=S_{1}^{R}=S_{BG}$ and $S_\alpha^R$ ($\forall \alpha$) are extensive, whereas the nonadditive entropy $S_q$ ($q \ne 1$) is nonextensive. In such case we have
\begin{equation}
s=S\left( 1\right) ,  \label{ss1}
\end{equation}%
where $S\left( 1\right) $ is the entropy of one element when it is
considered as isolated. 
Furthermore, for short-range
interacting systems, \textit{i.e.}, whose elements are only asymptotically independent  (%
\textit{e.g.}, air molecules at normal conditions),  i.e., strict independence is now violated, $S_{BG}$ and $S_\alpha^R$ still are extensive, whereas $S_q$ ($q\ne $) still is nonextensive.  
For such systems, 
$s_{BG}\equiv \lim_{N \to\infty} S_{BG}(N)/N \neq S\left(
1\right) $. Conversely, for subsystems of systems exhibiting long-range correlations~\cite{CarusoTsallis2007}, 
$ S_{BG}$ and $S_{\alpha }^{R}$ are  \textit{nonextensive}, \textit{whereas }$%
S_{q}$ \textit{can be extensive for an appropriate value of the entropic
index }$q\neq 1$.
This class of systems has been coined as $q$-\emph{describable} \cite{EurophysicsNews}.

\subsection{Optimising $S_{q}$}

Let us consider the continuous version of the nonadditive entropy $S_{q}$, 
\textit{i.e.}, 
\begin{equation}
S_{q}=k\frac{1-\int \left[ p\left( X\right) \right] ^{q}\ dX}{1-q}.
\label{sq-cont}
\end{equation}%
The natural constraints in the maximisation of (\ref{sq-cont}) are
(hereinafter $k=1$), $\int p\left( X\right) \ dX=1\,$, corresponding to
normalisation, and 
\begin{equation}
\int X\frac{\ \left[ p\left( X\right) \right] ^{q}}{\int \left[ p\left(
X\right) \right] ^{q}dX}\ dX\equiv \left\langle X\right\rangle _{q}=\bar{\mu}%
_{q}\,,
\end{equation}%
\begin{equation}
\int \left( X-\bar{\mu}_{q}\right) ^{2}\frac{\ \left[ p\left( X\right) %
\right] ^{q}}{\int \left[ p\left( X\right) \right] ^{q}dX}\ dX\equiv
\left\langle \left( X-\bar{\mu}_{q}\right) ^{2}\right\rangle _{q}=\bar{\sigma%
}_{q}^{2}\,,  \label{variance}
\end{equation}%
corresponding to the $q$-\textit{generalised} mean and variance of $X$,
respectively.

From the variational problem we obtain 
\begin{equation}
\mathcal{G}_{q}\left( X\right) =\mathcal{A}_{q}\left[ 1+\left( q-1\right) 
\mathcal{B}_{q}\left( X-\bar{\mu}_{q}\right) ^{2}\right] ^{\frac{1}{1-q}%
},\qquad \left( q<3\right) ,  \label{pq-1}
\end{equation}%
(if the quantity within brackets is nonnegative, and zero otherwise) where, 
\begin{equation}
\mathcal{A}_{q}=\left\{ 
\begin{array}{ccc}
\frac{\Gamma \left[ \frac{5-3q}{2-2q}\right] }{\Gamma \left[ \frac{2-q}{1-q}%
\right] }\sqrt{\frac{1-q}{\pi }\mathcal{B}_{q}} & \Leftarrow  & q<1 \\ 
\sqrt{\frac{\mathcal{B}_{q}}{\pi }} & \Leftarrow  & q=1 \\ 
\frac{\Gamma \left[ \frac{1}{q-1}\right] }{\Gamma \left[ \frac{3-q}{2q-2}%
\right] }\sqrt{\frac{q-1}{\pi }\mathcal{B}_{q}} & \Leftarrow  & q>1%
\end{array}%
\right. ,
\end{equation}%
and $\mathcal{B}_{q}=\left[ \left( 3-q\right) \,\bar{\sigma}_{q}^{2}\right]
^{-1}$. Standard and generalised variances, $\bar{\sigma}_{q}^{2}$ and $\bar{%
\sigma}^{2}$ are related through $\bar{\sigma}_{q}^{2}=\bar{\sigma}^{2}\frac{5-3q%
}{3-q}$, for $q<\frac{5}{3}$ .

Defining the $q$-\textit{exponential} function \footnote{%
Other generalisations for the exponential function can be found at Ref.~\cite%
{exp-gen}.}\ as 
\begin{equation}
e_{q}^{x}\equiv\left[ 1+\left( 1-q\right) \,x\right] ^{\frac{1}{1-q}%
}\qquad\left( e_{1}^{x}\equiv e^{x}\right) ,  \label{q-exp}
\end{equation}
($e_{q}^{x}=0$ if $1+(1-q)x\leq0$) we can rewrite PDF~(\ref{pq-1}) as 
\begin{equation}
\mathcal{G}_{q}\left( x\right) \equiv \mathcal{A}_{q}\,e_{q}^{-\mathcal{B}%
_{q}\left( x-\bar{\mu}_{q}\right) ^{2}},  \label{pq}
\end{equation}
hereon referred to as $q$-\emph{Gaussian}. The inverse function of the $q$%
-exponential, the $q$-logarithm, is $\ln_{q}\left( x\right) \equiv \frac{%
x^{1-q}-1}{1-q}\,\,(x>0)$.

For $q=\frac{3+m}{1+m}$, the $q$-Gaussian recovers Student's $t$%
-distribution with $m$ degrees of freedom ($m=1,2,3,\ldots $) and finite
moment up to order $m$. So, for $q>1,$ PDF~(\ref{pq}) presents an asymptotic 
\textit{power-law} behaviour. Complementarily, if $q=\frac{n-4}{n-2}$ with $%
n=3,4,5,\ldots $, $p\left( x\right) $ recovers the $r$-distribution with $n$
degrees of freedom. Consistently, for $q<1$, $p\left( x\right) $ has a 
\textit{compact support} defined by the condition $\left\vert x-%
\bar{\mu}_{q}\right\vert \leq \sqrt{\frac{3-q}{1-q}\,\bar{\sigma}_{q}^{2}}$ .

\subsection{$q$-calculus}

The nonadditivity property of $S_{q}$, assuming \textit{for independent systems} $A$ and $B$, 
\begin{equation}
\frac{S_{q}\left( A+B\right) }{k}=\frac{S_{q}\left( A\right) }{k}+\frac{%
S_{q}\left( B\right) }{k}+\left( 1-q\right) \frac{S_{q}\left( A\right) }{k}%
\frac{S_{q}\left( B\right) }{k},
\end{equation}%
has inspired the introduction of a new algebra~\cite{nivanen,borges}
composed by $q$\emph{-sum}, $x\oplus _{q}y\equiv x+y+\left( 1-q\right) x\,y$%
, and the $q$\emph{-product} 
\begin{equation}
x\otimes _{q}y\equiv \left[ x^{1-q}+y^{1-q}-1\right] ^{\frac{1}{1-q}}.
\label{q-multi}
\end{equation}%
The corresponding inverse operations are the $q$\emph{-difference}, $%
x\ominus _{q}y$, and the $q$\emph{-division},\emph{\ }$x\oslash _{q}y$, such
that, $\left( x\otimes _{q}y\right) \oslash _{q}y=x$. The $q$-product can be
written by using the basic function of nonextensive formalism, the $q$%
-exponential, and its inverse, the $q$-logarithm. Hence, $x\otimes
_{q}y\equiv \exp _{q}\left[ \ln _{q}\,x+\ln _{q}\,y\right] $, which for $%
q\rightarrow 1$, recovers the usual property $\ln \left( x\times y\right)
=\ln \,x+\ln \,y$ ($x,y>0$), where $x\times y\equiv x\otimes _{1}y$. Since $%
\exp _{q}\left[ x\right] $ is a non-negative function, the $q$-product must
be restricted to values of $x$ and $y$ that respect condition%
\begin{equation}
\left\vert x\right\vert ^{1-q}+\left\vert y\right\vert ^{1-q}-1\geq 0
\label{cond-q-prod}
\end{equation}%
We can enlarge the domain of the $q$-product to negative values of 
$x$ and $y$ by writing it as 
\begin{equation}
x\otimes _{q}y\equiv \mathrm{\ sign}\left( x\,y\right) \exp _{q}\left[ \ln
_{q}\,\left\vert x\right\vert +\ln _{q}\,\left\vert y\right\vert \right] .
\label{q-product-new}
\end{equation}%
We list now a set of properties of the $q$-product:

\begin{enumerate}
\item $x\otimes _{1}y=x\ y$ ;

\item $x\otimes _{q}y=y\otimes _{q}x$ ;

\item $\left( x\otimes _{q}y\right) \otimes _{q}z=x\otimes _{q}\left(
y\otimes _{q}z\right) =    x\otimes _{q}
y\otimes _{q}z    =\left[ x^{1-q}+y^{1-q}+z^{1-q}-3\right] ^{\frac{1}{1-q}}$ ;

\item $\left( x\otimes _{q}1\right) =x$ ;

\item $\ln _{q}\left[ x\otimes _{q}y\right] \equiv \ln _{q}\,x+\ln _{q}\,y$ ;

\item $\ln _{q}\left( x\,y\right) =\ln _{q}\left( x\right) +\ln _{q}\left(
y\right) +\left( 1-q\right) \ln _{q}\left( x\right) \ln _{q}\left( y\right) $%
;

\item $\left( x\otimes _{q}y\right) ^{-1}=x^{-1}\otimes _{2-q}y^{-1}$;

\item $\left( x\otimes _{q}0\right) =\left\{ 
\begin{array}{ccc}
0 &  & \mathrm{if\ }\left( q\geq 1\ \mathrm{and\ }x\geq 0\right) \mathrm{or\
if\ }\left( q<1\ \mathrm{and\ }0\leq x\leq 1\right)\,, \\ 
&  &  \\ 
\left( x^{1-q}-1\right) ^{\frac{1}{1-q}} &  & \mathrm{if \,\,q<1 \, and \, x>1} .%
\end{array}%
\right. $
\end{enumerate}

\begin{figure}[tbh]
\centering
\includegraphics[width=0.75\columnwidth,angle=0]{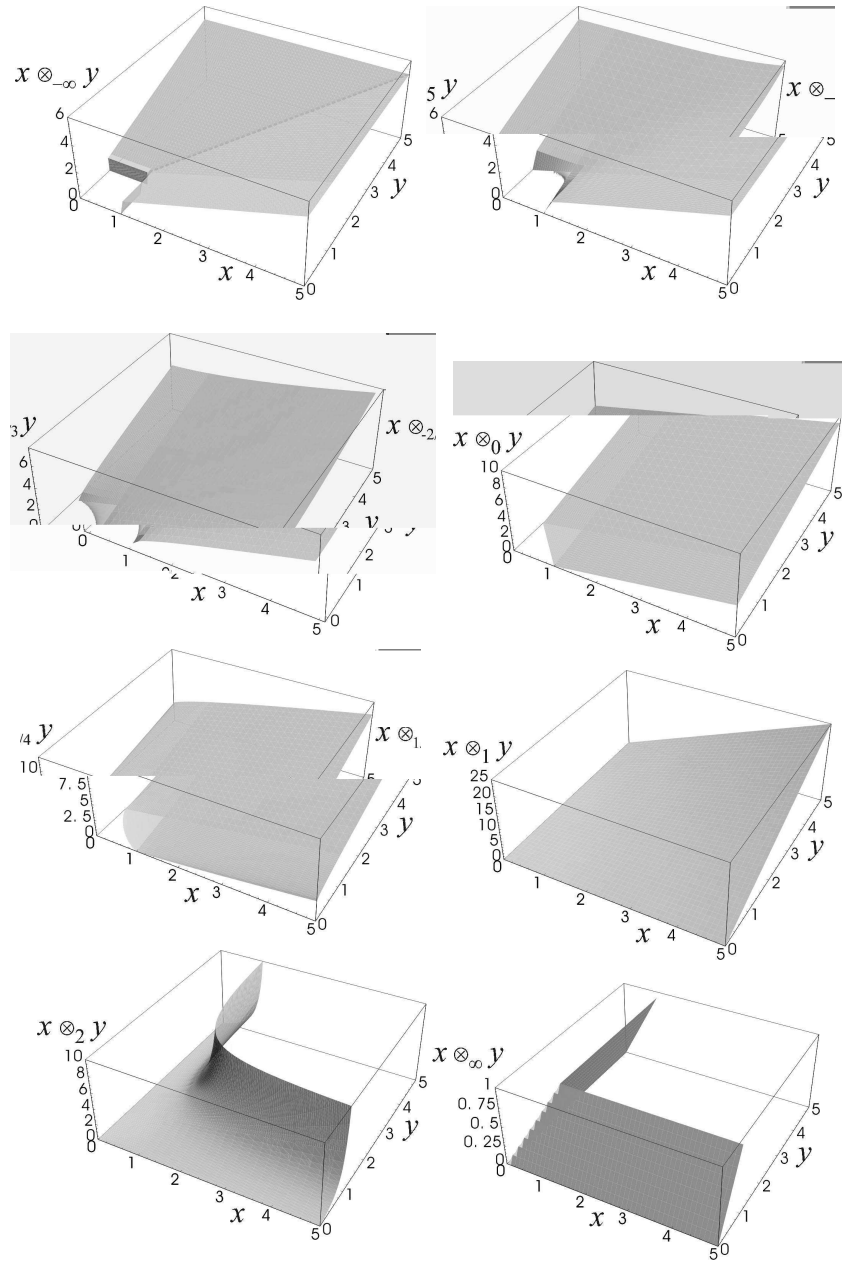} %
\caption{Representation of the $q$-product, Eq.~(\protect\ref{q-product-new}%
), for $q=-\infty$, $-5$, $-2/3$, $0$, $1/4$, $1$, $2$, $\infty$. As it is
visible, the squared region $\left\{ 0\leq x\leq1,0\leq y\leq1\right\} $ is
gradually integrated into the nontrivial domain as $q$ increases up to $q=1$. From this
value on, a new prohibited region appears, but this time coming from large
values of $\left( \vert x \vert, \vert y \vert\right) $. This region reaches
its maximum when $q=\infty$. In this case, the domain is composed by a
horizontal and vertical strip of width $1$.}
\label{qproduct}
\end{figure}

For special values of $q$, \textit{e.g.}, $q=1/2$, the argument of the $q$-product can
attain nonpositive values, specifically at points for which $\left\vert x\right\vert
^{1-q}+\left\vert y\right\vert ^{1-q}-1<0$. In these cases, and consistently with 
the cut-off for the $q$-exponential we have set $x\otimes _{q}y=0$. With regard to the $q$-product domain, and restricting our analysis of Eq. (\ref%
{cond-q-prod}) to $x,y>0$, we observe that for $q\rightarrow -\infty $ the
region $\left\{ 0\leq x\leq 1,0\leq y\leq 1\right\} $ leads to a vanishing $q$-product. As the
value of $q$ increases, the forbidden region decreases its area, and when $%
q=0$ we have the limiting line given by $x+y=1$, for which $x\otimes _{0}y=0$%
. Only for $q=1$, the whole set of real values of $x$ and $y$ has a defined
value for the $q$-product. For $q>1$,\ condition (\ref{cond-q-prod}) yields
a curve, $\left\vert x\right\vert ^{1-q}+\left\vert y\right\vert
^{1-q}=1 $, at which the $q$-product diverges. This undefined region
increases as $q$ goes to infinity. At the $q\to\infty$ limit, the $q$-product
is only defined in $\left\{ x>1,y\leq 1\right\} \cup \left\{ 0\leq x\leq
1,0\leq y\leq 1\right\} \cup \left\{ x\leq 1,y>1\right\} $. This entire
scenario is depicted on the panels of Fig.~\ref{qproduct}. The profiles presented by $x\otimes _{\infty }y$ and $%
x\otimes _{-\infty }y$ illustrate the above property (8). To illustrate the $q$-product in another simple form, we show, in Fig.~\ref%
{q-prod-x-x}, a representation of $x \otimes_{q}x$ \ for typical values of $%
q $. 

\begin{figure}[tbh]
\includegraphics[width=0.4\columnwidth,angle=0]{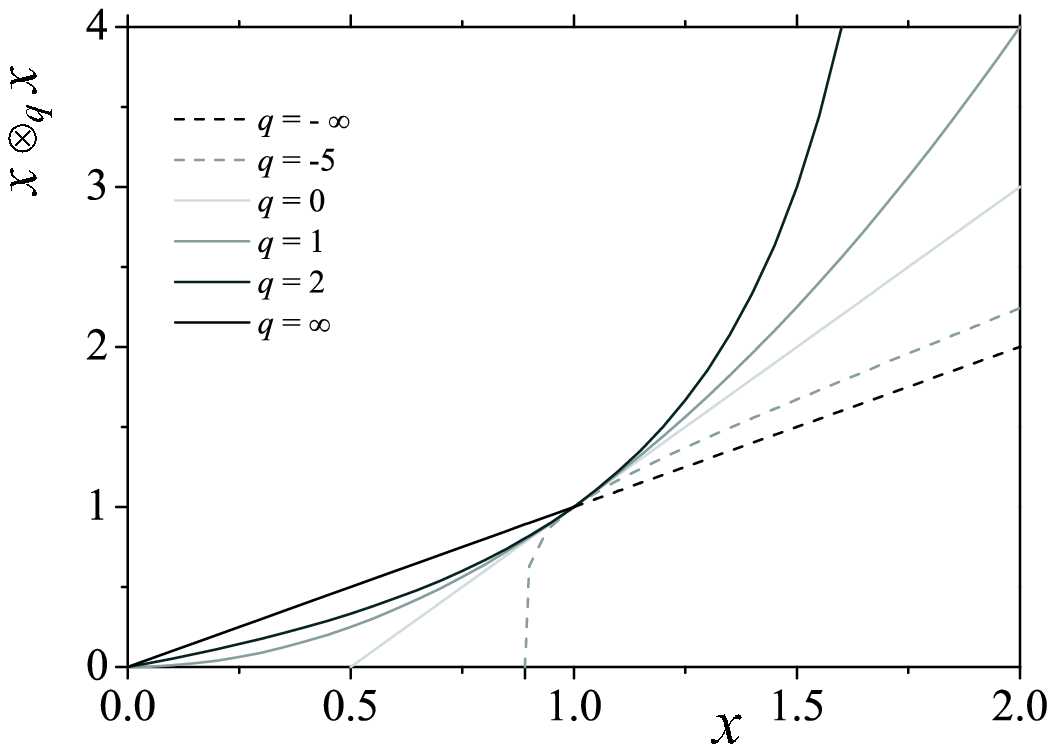} 
\caption{Representation of the $q$-product, $x\otimes_{q}x$ for $q=-\infty$, 
$-5$, $0$, $1$, $2$, $\infty$. Excluding $q=1$, there is a special value $%
x^{\ast}=2^{1/\left( q-1\right) }$, for which $q<1$ represents the lower
bound [in figure $x^{\ast}\left( q=-5\right) =2^{-1/6}\simeq0.89089$ and $%
x^{\ast}\left( q=0\right) =1/2$], and for $q>1$ the upper bound [in figure $%
x^{\ast}\left( q=2\right) =2$]. For $q=\pm\infty$, $x\otimes_{q}x$ lies on
the diagonal of bisection, but following the lower and upper limits
mentioned above.}
\label{q-prod-x-x}
\end{figure}

\section{L\'{e}vy distributions}

In the context of the CLT for independent variables, apart from the Gaussian
distribution, $\mathcal{G}\left( X\right) $, another stable distribution
plays a key role, the $\alpha$-L\'{e}vy distribution, $L_{\alpha}^{B}\left(
X\right) $ ($0<\alpha<2$). If $\mathcal{G}\left( X\right) $ is characterised
by its fast decay, $L_{\alpha}^{B}\left( X\right) $ is characterised by its
`fat' tails, since it allows both small and large values of $X$ to be effectively measurable. Widely applied in several areas, $%
L_{\alpha}^{B}\left( X\right) $ is defined through its Fourier Transform 
\cite{levy},%
\begin{equation}
\begin{array}{ccc}
\hat{L}_{\alpha}^{B}\left( k\right) & = & \exp\left[ -a\,\left\vert
k\right\vert ^{\alpha}\left\{ 1+i\,B\tan\left( \alpha\frac{\pi}{2}\right) 
\frac{k}{\,\left\vert k\right\vert }\right\} \,\right] ,\qquad\left(
\alpha\neq1\right) ,%
\end{array}
\label{eq-levy}
\end{equation}
(where $\alpha$ is the \emph{L\'{e}vy exponent,} and $B$ represents the 
\emph{asymmetry parameter}). By this we mean that L\'{e}vy distributions have
no\ analytical form in $X$, excepting for special values of $\alpha=1$. For $B=0$, $L_{\alpha}^{0}\equiv L_{\alpha}$, the distribution is
symmetric. Regarding $\alpha$ values, one can verify that $L_{\frac{1}{2}%
}\left( x\right) $ corresponds to the \textit{L\'{e}vy-Smirnov}
distribution, and that $L_{1}\left( X\right) $ is the \textit{Cauchy} or 
\textit{Lorentz} distribution ($L_{1}\left( X\right) $ coincides with the  $\mathcal{G%
}_{2}\left( X\right) $ distribution). For $\alpha=2$, $\hat{L}%
_{\alpha}\left( k\right) $ has a Gaussian form, thus the corresponding
distribution is\ a Gaussian.

\begin{figure}[tbh]
\includegraphics[width=0.4\columnwidth,angle=0]{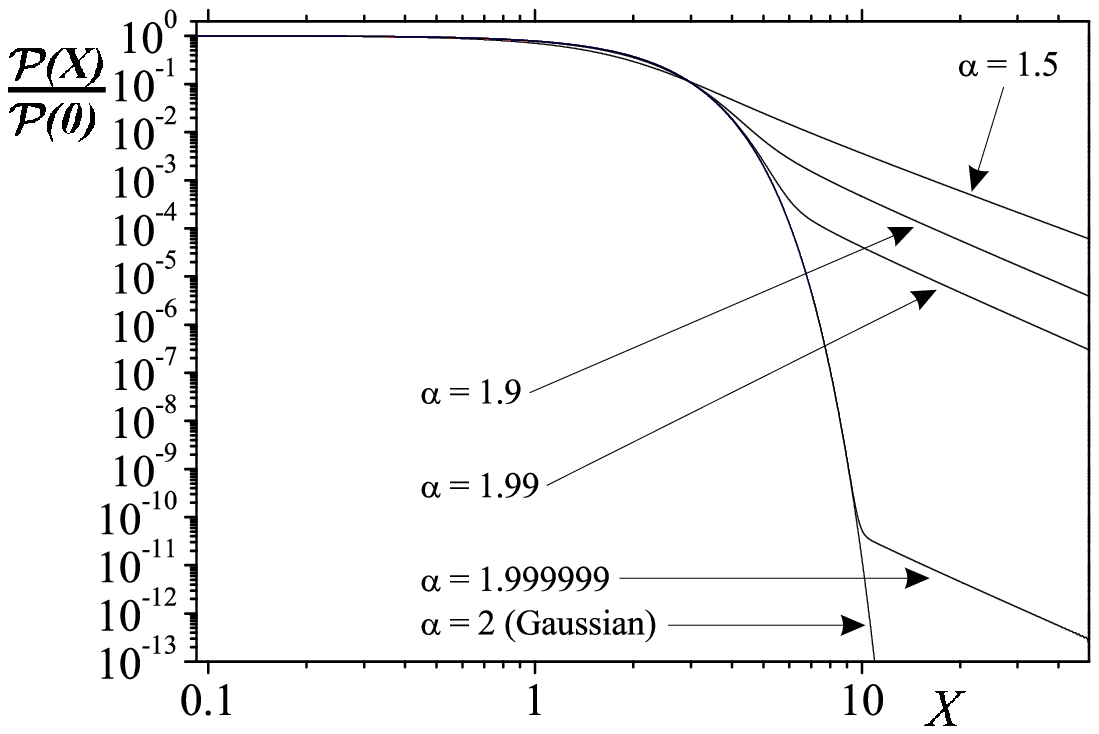}
\includegraphics[width=0.32\columnwidth,angle=0]{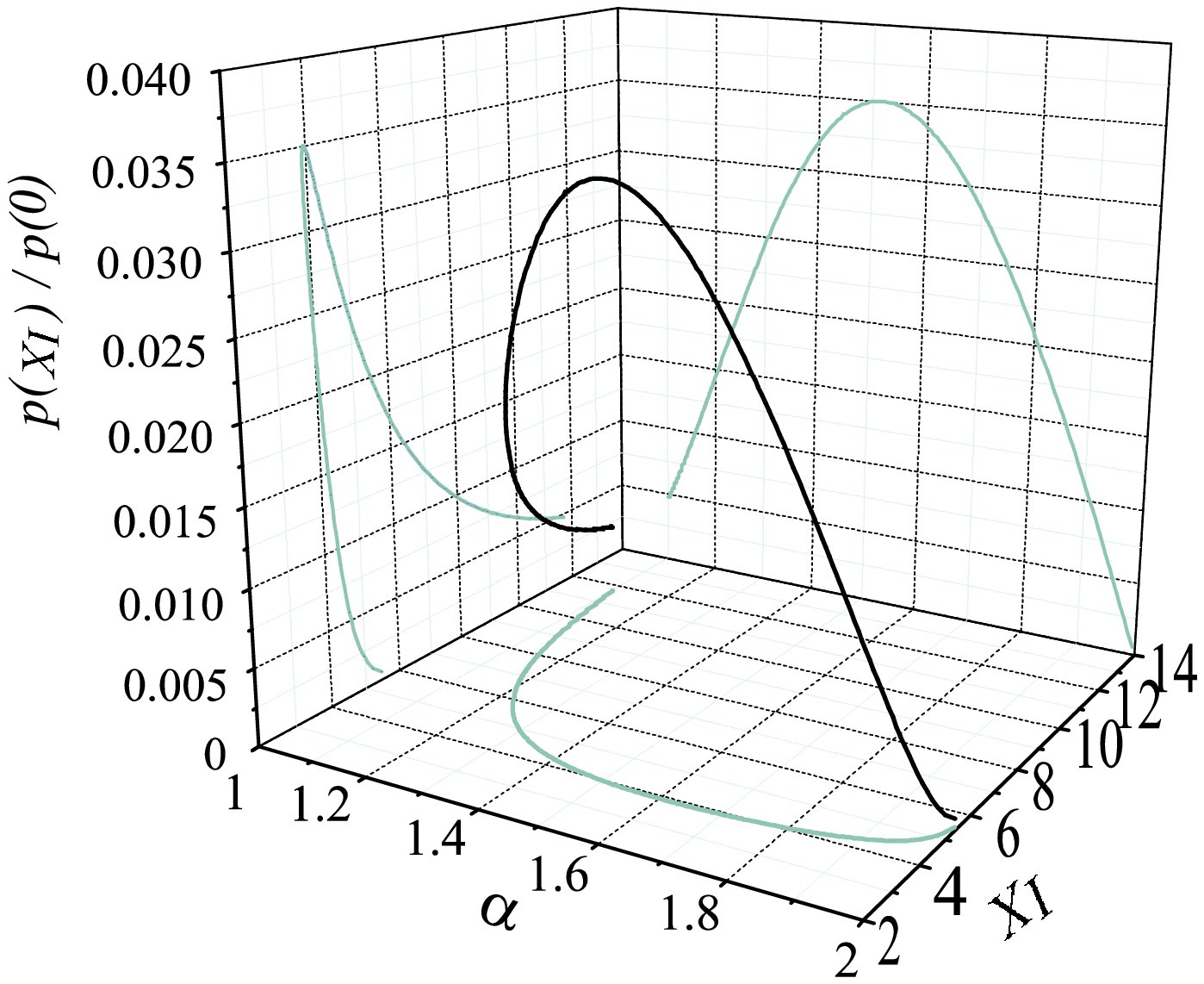} 
\caption{{\it Left panel}: Gaussian and $\protect\alpha$-stable L\'{e}vy distributions for $%
\protect\alpha$ approaching $2$ in Eq.~(\protect\ref{eq-levy}) with $a = 1$ and
$B = 0$. As referred in the text for values of $\protect\alpha$ closer to $2
$, L\'{e}vy distribution becomes almost equal to a Gaussian up to some
critical value for which the power law behaviour emerges. {\it Right panel}: Locus of the inflexion point of the $\protect\alpha$-stable L\'{e}vy distributions, Eq.~(%
\protect\ref{eq-levy}), with $a = 1$ and $B = 0$. Contrarily to what happens
with $\mathcal{G}_{q}\left( X\right) $, when L\'{e}vy distributions are
represented in a $\log$-$\log$ scale, they exhibit an inflexion point which
goes to infinity as $\protect\alpha\rightarrow1$ (Cauchy-Lorentz
distribution $\mathcal{G}_{2}\left( X\right) $) and $\protect\alpha%
\rightarrow2$ (Gaussian distribution $\mathcal{G}\left( X\right) $) too. We
also show the projections onto the planes $\frac{p\left( X_{I} \right) }{p\left(
0\right) }-X_{I}$, $\frac{p\left( X_{I} \right) }{p\left( 0\right) }-\protect%
\alpha$, and $\protect\alpha-X_{I}$. }
\label{locus}
\end{figure}

Carrying out the inverse Fourier Transform on $\tilde{L}_{\alpha }\left(
k\right) $, $L_{\alpha }\left( X\right) =\frac{1}{2\,\pi }\int_{-\infty
}^{+\infty }e^{-i\,k\,X}\,\hat{L}_{\alpha }\left( k\right) \,dk$, we can
straightforwardly evaluate the limit for small $X$, $L_{\alpha }\left(
X\right) \approx \left( \pi \,\alpha \,a^{a/\alpha }\right) ^{-1}\Gamma %
\left[ \frac{1}{\alpha }\right] $, and the limit for large $X$,%
\begin{equation}
L_{\alpha }\left( X\right) \sim \frac{a\,\alpha }{\pi }\frac{\Gamma \left[
\alpha \right] \sin \left[ \frac{\pi \,\alpha }{2}\right] }{\left\vert
X\right\vert ^{1+\alpha }},\qquad X\rightarrow \infty .
\label{limit-infinito}
\end{equation}%
From condition, $0<\alpha <2$, it is easy to prove that variables associated
with a L\'{e}vy distribution do not have a finite second-order moment, just
like $\mathcal{G}_{q}\left( X\right) $ with $\frac{5}{3}\leq q<3$ (see \cite
{ct-prato} and references therein). In spite of the fact that we can write two distributions, $%
\mathcal{G}_{q}\left( X\right) $ and $L_{\alpha }\left( X\right) $, which
present the same asymptotic power-law decay (with $\alpha =\frac{3-q}{q-1}$%
), there are interesting differences between them. The first one is that, as
we shall see later on, $L_{\alpha }\left( X\right) $ together with $\mathcal{%
G}\left( X\right) $ are the only two stable functional forms whenever we
convolute \textit{independent} variables \cite{araujo}. Therefore,
distribution $\mathcal{G}_{q}\left( X\right) $ ($q\neq 1\,,2$) is not stable
for this case. The other point concerns their representation in a $\log
-\log $ scale. Contrarily to what happens in  $\log -\log $ representations of $\mathcal{G}_{q}\left( X\right) 
$, an
inflexion point exists at $X_{I}$ for $L_{\alpha }\left( X\right) $ if $1<\alpha
<2$, see Fig.~\ref{locus}. Since in many cases the numerical adjustment of
several experimental/computational probability density functions for either a L%
\'{e}vy or a $q$-Gaussian distribution seems to be plausible, the presence
of an inflexion point might be used as an extra criterion to conclude which
one is the most adequate. This has clear implications on phenomena modelling.

\section{Central limit theorems for independent variables}

\subsection{Variables with finite variance}

\label{clt-gaussiano}

Let us consider a sequence $X_{1},X_{2},\ldots,X_{N}$ of random variables
which are defined on the same probability space, share the same probability
density function, $p\left( X\right) $, with $\sigma <\infty$, and are
independent in the sense that the joint probability density function of any
two $X_{i}$ and $X_{j}$, $P\left( X_{i},X_{j}\right) $, is just $p\left(
X_{i}\right) p\left( X_{j}\right) $.

Hence \cite{lindeberg,feller}, a new variable%
\begin{equation}
Y=\frac{X_{1}+X_{2}+\ldots +X_{N}}{N},  \label{soma-gaussiana}
\end{equation}%
with raw moments, $\left\langle Y^{n}\right\rangle \equiv \left\langle \frac{%
1}{N^{n}}\left( \sum\limits_{j=1}^{N}X_{j}\right) ^{n}\right\rangle $, has
its probability density function given by the convolution of $N$ probability
density functions, or, since variables are independent, by the Fourier
Transform,%
\begin{equation}
\begin{array}{ccc}
\mathcal{F}\left[ P\left( Y\right) \right] \left( k\right)  & = & \frac{1}{%
2\pi }\left\{ \int e^{i\,k\,\frac{X}{N}}p\left( X\right) \,dX\right\} ^{N}=%
\frac{1}{2\pi }\left\{ \int \sum\limits_{n=0}^{\infty }\frac{\left(
ik\right) ^{n}}{n!}\frac{\left\langle X^{n}\right\rangle }{N}\,dX\right\}
^{N} \\ 
& = & \frac{1}{2\pi }\exp \left[ N\,\ln \left[ 1+ik\frac{\left\langle
X\right\rangle }{N}-\frac{1}{2}k^{2}\frac{\left\langle X^{2}\right\rangle }{%
N^{2}}+O\left( N^{-3}\right) \right] \right] .%
\end{array}%
\end{equation}%
Since $\ln \left( 1+x\right) =x-\frac{x^{2}}{2}+\frac{x^{3}}{3}+\ldots $,
expanding up to second order in $X$ we asymptotically have, in the limit $
N\rightarrow \infty $ ,
\begin{equation}
\begin{array}{ccc}
\mathcal{F}\left[ P\left( Y\right) \right] \left( k\right)  & \approx  & 
\frac{1}{2\pi }\exp \left[ N\left\{ +i\,k\,\frac{\left\langle X\right\rangle 
}{N}-\frac{k^{2}}{2N^{2}}\left( \left\langle X^{2}\right\rangle
-\left\langle X\right\rangle ^{2}\right) \right\} \right] .%
\end{array}%
\end{equation}%
Defining $\mu _{X}\equiv \left\langle X\right\rangle ,$ as the mean value,
and $\sigma _{X}^{2}\equiv \left\langle X^{2}\right\rangle -\left\langle
X\right\rangle ^{2},$ as the standard deviation we have 
\begin{equation}
\begin{array}{ccc}
\mathcal{F}\left[ P\left( Y\right) \right] \left( k\right)  & \approx  & 
\frac{1}{2\pi }\exp \left[ -i\,k\,\mu _{X}-\frac{k^{2}}{2N}\sigma _{X}^{2}%
\right] 
\end{array}%
.
\end{equation}%
Performing the Inverse Fourier Transform for $\mathcal{F}\left[ P\left(
Y\right) \right] \left( k\right) $, we obtain the distribution $P\left(
Y\right) ,$%
\begin{equation}
\begin{array}{ccc}
P\left( Y\right)  & = & \int e^{-ikY}\mathcal{F}\left[ P\left( Y\right) %
\right] \left( k\right) \,dk%
\end{array}%
,  \label{py-definicao}
\end{equation}%
which yields,%
\begin{equation}
\begin{array}{ccc}
P\left( Y\right)  & \approx  & \frac{1}{\sqrt{2\pi }\sqrt{\frac{\sigma
_{X}^{2}}{N}}}e^{-\left( Y-\mu _{X}\right) ^{2}/(2\frac{\sigma _{X}^{2}}{N})}%
\end{array}%
\end{equation}%
Remembering that, from Eq.~(\ref{soma-gaussiana}) $\mu _{Y}=\mu _{X}$ and $%
\sigma _{Y}=\sigma _{X}\,N^{-1/2}$ we finally get%
\begin{equation}
\begin{array}{ccc}
P\left( Y\right)  & = & \frac{1}{\sqrt{2\pi }\sigma _{Y}}e^{-\left( Y-\mu
_{Y}\right) ^{2}/(2\,\sigma _{Y}^{2})}%
\end{array}%
,
\end{equation}%
which is a Gaussian. It is also easy to verify that $P\left( Z=\frac{Y}{%
\sqrt{N}}\right) =\sqrt{N}P\left( Y\right) $, \textit{i.e.}, the Gaussian
distribution scales as $N^{-\frac{1}{2}}$.

\subsubsection{Example: The convolution of independent $q$-variables}

\label{clt-levy}

\begin{figure}[tbh]
\includegraphics[width=0.5\columnwidth,angle=0]{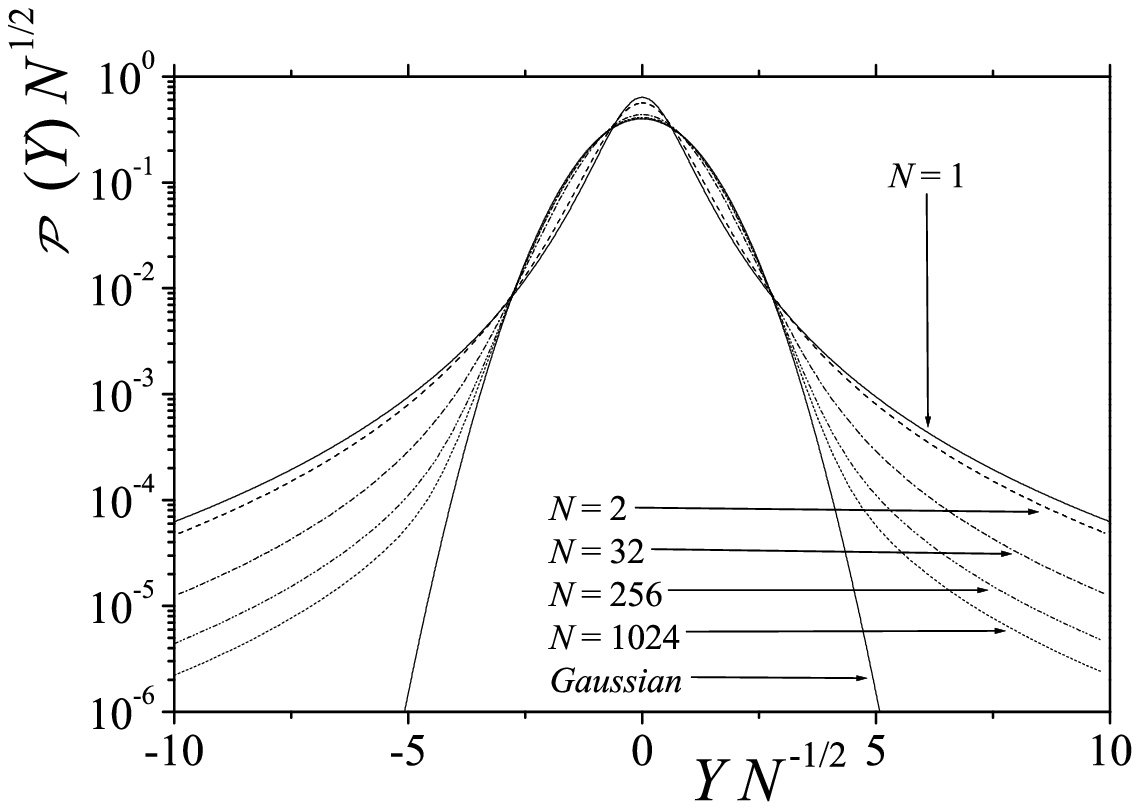} %
\includegraphics[width=0.5\columnwidth,angle=0]{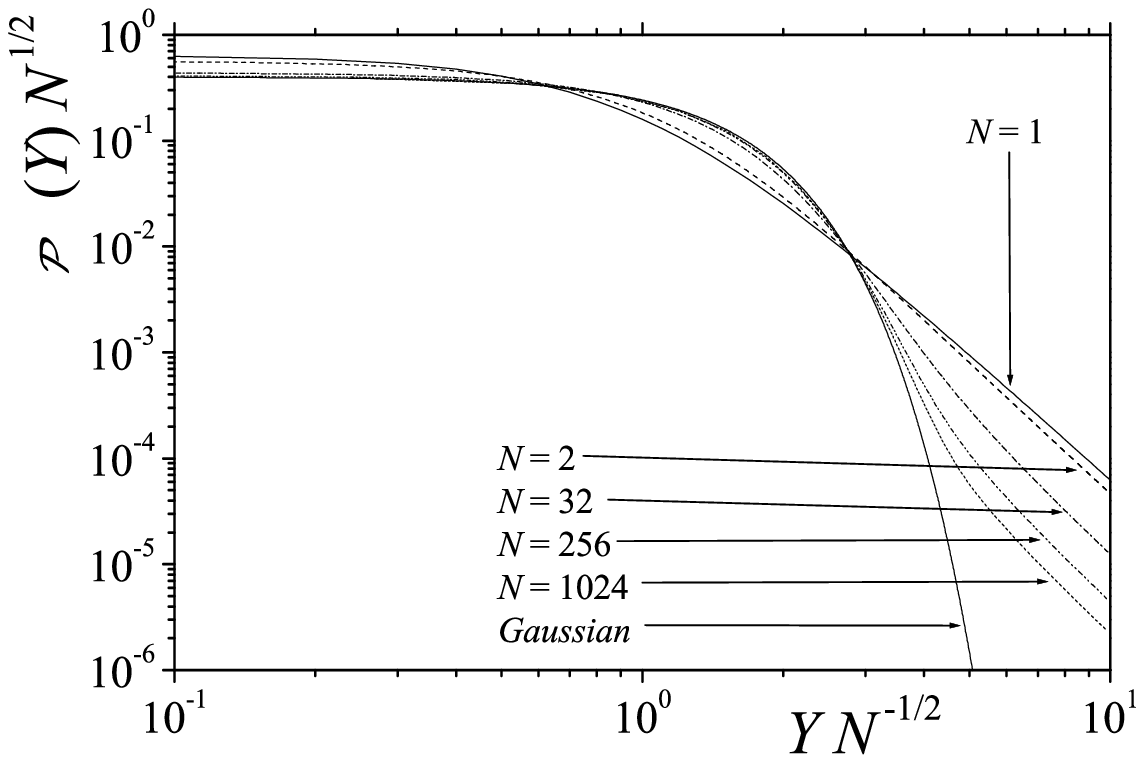} 
\caption{Both panels represent probability density function $\mathcal{P}%
\left( Y\right) $ \textit{vs.} $Y$ (properly scaled) in  log-linear (left)
and $\log$-$\log$ (right) scales, where $Y$ represents the sum of $N$
independent variables $X$ having a $\mathcal{G}_{\frac{3}{2}}\left( X\right) 
$ distribution. Since variables are independent and its variance is finite, $%
\mathcal{P}\left( Y\right) $ converges to a Gaussian as it is visible. It is
also visible in the log-linear representation that, although the central
part of the distribution approaches a Gaussian, the power-law decay subsists
even for large $N$ as it is depicted in $\log$-$\log$ representation. }
\label{q-gauss15}
\end{figure}

Let us assume that the probability density function $p\left( X\right) $ is a 
$q$-Gaussian distribution, $\mathcal{G}_{q}(X)$, which, as a simple illustration, has $q=\frac{3}{2}$, and $\sigma =1< \infty $ \footnote{%
The value $q=\frac{3}{2}$ appears in a wide range of phenomena which goes
from long-range hamiltonian systems \cite%
{latora-rapisarda-tsallis,tsallis-angra,celia-goyo} to economical systems 
\cite{obt}.}. In this case, the Fourier Transform for $\mathcal{G}_{\frac{3}{%
2}}\left( X\right) $ is $\mathcal{F}\left[ \mathcal{G}_{\frac{3}{2}}\left(
X\right) \right] \left( k\right) =\left( 1+\left\vert k\right\vert \right)
\exp \left[ -\left\vert k\right\vert \right] $. The distribution, $\mathcal{P%
}\left( Y\right) $, where $Y=X_{1}+X_{2}+\ldots +X_{N}$, is given by%
\begin{equation}
\begin{array}{ccc}
\mathcal{P}\left( Y\right)  & = & \frac{1}{2\pi }\int \exp \left[ -i\,k\,Y%
\right] \left\{ \mathcal{F}\left[ \mathcal{G}_{\frac{3}{2}}\left( X\right) %
\right] \left( k\right) \right\} ^{N}dk%
\end{array}%
.
\end{equation}%
Expanding $\left\{ \mathcal{F}\left[ \mathcal{G}_{\frac{3}{2}}\left(
X\right) \right] \left( k\right) \right\} ^{N}$ around $k=0$ we obtain, $%
\left\{ \mathcal{F}\left[ \mathcal{G}_{\frac{3}{2}}\left( X\right) \right]
\left( k\right) \right\} ^{N}\simeq 1-\frac{1}{2}Nk^{2}+\frac{1}{3}%
N\left\vert k\right\vert ^{3}$. From the CLT, we observe that distribution $%
\mathcal{P}\left( Y\right) $, for large $N$, is well described by a
Gaussian, $\mathcal{G}\left( Y\right) \approx \frac{1}{\sqrt{2\,\pi \,N}}%
\exp \left[ -\frac{Y^{2}}{2N}\right] $, at its central region (for $N$
large). Because of singularity $\left\vert k\right\vert ^{3}$, associated
with the divergence in $n$-order statistical moments ($n\geq 3$), we have
the remaining distribution described by a power-law, $\mathcal{P}\left(
Y\right) \sim \frac{2N}{\pi }Y^{4}$, for large $Y$. This specific behaviour
is depicted in Fig.~\ref{q-gauss15}. Therein we observe a crossover from
Gaussian to power law at $Y\,N^{-1/2}$ of order $\sqrt{\ln N}$, which tends
to infinity.

\subsection{Variables with infinite variance}

Figure~\ref{q-gauss15} is similar to what is presented in Fig.~\ref{locus} (left panel). As
it is visible there, as $\alpha $ goes to $2$, the
distribution $L_{\alpha }\left( X\right) $ nearly collapses onto the
Gaussian distribution up to some critical value $X^{\ast }$. Beyond that
point, the asymptotic power-law character emerges and the distribution falls
as $\left\vert X\right\vert ^{-\alpha -1}$.

If we consider the sum, $Y=X_{1}+X_{2}+\ldots +X_{N}$, of $N$ random
variables, $X_{1},X_{2},\ldots ,X_{N}$, which share the same L\'{e}vy
distribution, $L_{\alpha }\left( X\right) $. The distribution $\mathcal{P}%
\left( Y\right) $ is then given by,%
\begin{equation}
\begin{array}{ccc}
\mathcal{P}\left( Y\right)  & = & \frac{1}{2\,\pi }\int_{-\infty }^{+\infty
}e^{-i\,k\,Y}\,\left\{ \tilde{L}_{\alpha }\left( k\right) \right\} ^{N}\,dk
\\ 
& = & \frac{1}{2\,\pi }\int_{-\infty }^{+\infty }\exp \left[
-i\,k\,Y-a\,N\,\left\vert k\right\vert ^{\alpha }\right] \,dk.%
\end{array}%
\end{equation}%
Introducing a new variable, $\omega =k\,N^{1/\alpha }$, we get $\mathcal{P}%
\left( Y\right) =N^{-1/\alpha }L_{\alpha }\left( \frac{Y}{N^{1/\alpha }}%
\right) $. Hence, for L\'{e}vy distributions we also have the scaling
property, but with exponent $1/\alpha $. This generalised version of the
Central Limit Theorem, originally due to Gnedenko-Kolmogorov \cite{gnedenko}%
, is applied to any distribution $p\left( x\right) $ which, in the limit $%
x\rightarrow \infty $, behaves as $\left\vert x\right\vert ^{-\mu -1}$ ($\mu
<2$). In other words, \emph{the probability density function of the sum of }$N$%
\emph{\ variables, each one with the same distribution }$p\left( x\right)
\sim \left\vert x\right\vert ^{-\alpha -1}$ ($0<\mu <2$)\emph{, converges, in
the $N\to\infty$ limit,} to a $\alpha $\emph{-stable L\'{e}vy
distribution}.

\section{Final Remarks}

\label{conclusion}

In this article we have reviewed central limit theorems for the sum of
independent random variables. As we have illustrated, in the absence of
correlations, the convolution of identical probability density functions
which maximise non-additive entropy $S_{q}$, behave in the same way as any
other distribution. In other words, when the entropic index $q<\frac{5}{3}$
the variance of $\mathcal{G}_{q}\left( X\right) $ is finite, hence the
convolution leads to a Gaussian distribution. On the other hand, \textit{%
i.e.}, when $q\geq \frac{5}{3}$ the variance diverges. As a consequence the
convolution leads to a $\alpha $-stable L\'{e}vy distribution with the same
asymptotic power-law decay of $\mathcal{G}_{q}\left( X\right) $ (the marginal case $q=5/3$ yields a Gaussian distribution, but with a logarithmic correction on the standard $x^2 \propto t$ scaling, i.e., it yields anomalous diffusion). We 
have also exhibited that there is an important difference between $q$-Gaussian and $%
\alpha $-stable L\'{e}vy distributions, namely the emergence of an
inflexion point on the latter type when they are represented in a $\log $-$\log $
scale. In the subsequent paper (Part II) we will show that the strong violation of the
independence condition introduces a drastic change in the probability space
attractor. Specifically, for a special class of correlations ($q$%
-{\it independence}), it is the $q$-Gaussian distribution which is a stable one.

\begin{theacknowledgments}
We are deeply thankful to E.P. Borges and R. Hilfer 
for fruitful discussions related to both Part I and Part II. Partial financial support from Pronex, CNPq,
Faperj (Brazilian agencies) and FCT/MCES (Portuguese agency) is acknowledged
as well.
\end{theacknowledgments}

\bibliographystyle{aipprocl} 


\end{document}